\documentclass[%
 aip,
 jcp,%
 amsmath,amssymb,
preprint,%
reprint,%
]{revtex4-1}

\usepackage{graphicx}
\usepackage{dcolumn}
\usepackage{bm}
\usepackage{xcolor}

\begin{document}

\author{Jakub K. Sowa}
\email{jakub.sowa@oxon.org}
\affiliation
{Department of Chemistry, Rice University, Houston, Texas 77005, USA}
\author{Danielle M. Cadena}
\affiliation
{Department of Chemistry, The University
of Texas at Austin, Austin, Texas 78712, USA}
\author{Arshad Mehmood}
\affiliation
{Department of Chemistry, Stony Brook University, Stony Brook, New York 11794, USA}
\affiliation{Institute of Advanced Computational Science, Stony Brook University, Stony Brook, New York 11794, USA}
\author{Benjamin G. Levine}
\affiliation
{Department of Chemistry, Stony Brook University, Stony Brook, New York 11794, USA}
\affiliation{Institute of Advanced Computational Science, Stony Brook University, Stony Brook, New York 11794, USA}
\author{Sean T. Roberts}
\email{roberts@cm.utexas.edu}
\affiliation
{Department of Chemistry, The University
of Texas at Austin, Austin, Texas 78712, USA}
\author{Peter J. Rossky}
\email{peter.rossky@rice.edu}
\affiliation
{Department of Chemistry, Rice University, Houston, Texas 77005, USA}

\title
  {IR Spectroscopy of Carboxylate-Passivated Semiconducting Nanocrystals: Simulation and Experiment}

\begin{abstract}
Surfaces of colloidal nanocrystals are frequently passivated with carboxylate ligands which exert significant effects on their optoelectronic properties and chemical stability. Experimentally, binding geometries of such ligands are typically investigated using vibrational spectroscopy, but the interpretation of the IR signal is usually not trivial. Here, using machine-learning (ML) algorithms trained on DFT data, we simulate an IR spectrum of a lead-rich PbS nanocrystal passivated with butyrate ligands. We obtain a good agreement with the experimental signal and demonstrate that the observed line shape stems from a very wide range of `tilted-bridge'-type geometries and does not indicate a coexistence of `bridging' and `chelating' binding modes as has been previously assumed. This work illustrates limitations of empirical spectrum assignment and demonstrates the effectiveness of ML-driven molecular dynamics simulations in reproducing IR spectra of nanoscopic systems. 
\end{abstract}

\maketitle
\section{Introduction}
Although quantum-confined colloidal nanocrystals (or quantum dots) possess size-dependent and therefore tunable properties,\cite{brus1984electron,alivisatos1996perspectives,cademartiri2006size,smith2010semiconductor} their optoelectronic as well as chemical characteristics can be strongly influenced by ligands passivating their surfaces.\cite{krause2015linking,calvin2022role} Consequently, various ligands have been used, for instance, to shift the positions of the nanocrystal band edges,\cite{kroupa2017tuning,brown2014energy} to enhance their optical absorbance,\cite{giansante2015darker} as well as to ensure the chemical stability of the quantum dots.\cite{choi2013steric,venettacci2019increasing} In recent years, organic ligands have also emerged as attractive energy or charge acceptors in nanocrystal-based optical and electronic devices.\cite{harris2016electronic,bender2018surface,cadena2022aggregation}

\begin{figure}[b]
    \centering
    \includegraphics{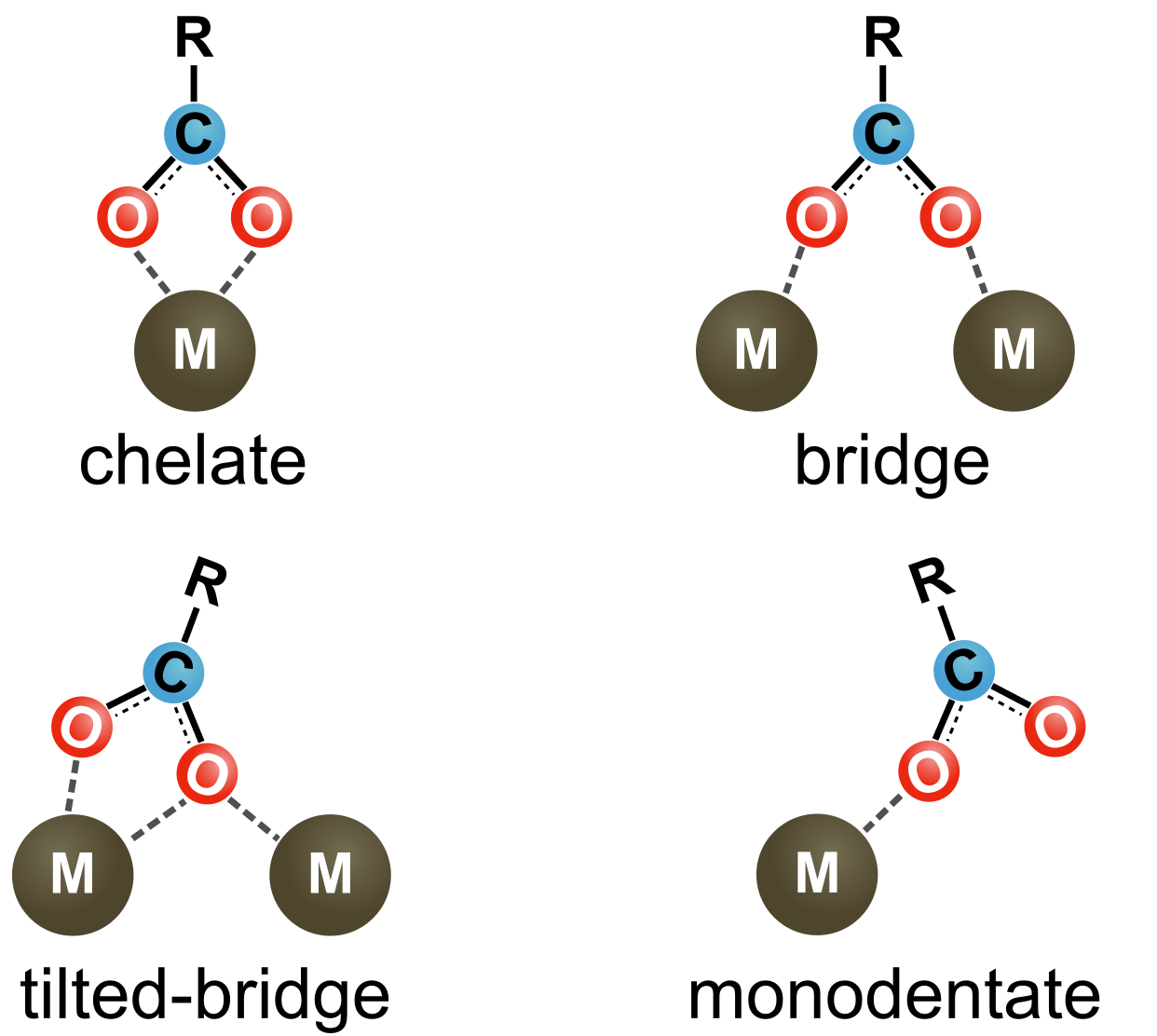}
    \caption{Schematic illustrations of the four binding modes of carboxylate ligands typically considered in literature.}
    \label{sketch}
\end{figure}

In practice, carboxylate-functionalized ligands are  frequently used both during nanocrystal synthesis and in post-synthesis surface modification.\cite{hines2003colloidal,choi2013steric,cadena2022aggregation,padgaonkar2021light} Such ligands can bind to the nanocrystal surface in different geometries of which four types (chelate, bridge, tilted-bridge, and monodentate) are usually identified depending on the number and position of metal-oxygen bonds,\cite{voznyy2011mobile,cass2013chemical,zhang2019identification} see Figure \ref{sketch}. These modes of binding are typically investigated using IR spectroscopy as the carboxylate-metal bonding is known to influence the carboxyl stretch frequencies. \cite{cass2013chemical,klein2023measuring,zhang2019identification,kennehan2020dynamic,peters2019sizing,peters2019hybrid,cros2010surface} 
In particular, the IR line shape as well as the splitting between the symmetric and antisymmetric C-O stretching bands, generally found in the region of approximately 1350-1650 cm$^{-1}$, are conventionally used to infer the existing binding geometries. A commonly used rule-of-thumb, derived from experimental studies of metal complexes,\cite{deacon1980relationships,zelevnak2007correlation} states that for chelating ligands the splitting between the symmetric and antisymmetric C-O stretch bands, $\Delta$, is usually around and below 100 cm$^{-1}$, for bridges this splitting is found between 100 and 150 cm$^{-1}$, while in monodentate ligands it exceeds 200 cm$^{-1}$. Nonetheless, given the presence of relatively broad and overlapping IR bands in experimental spectra of nanocrystals, an unambiguous interpretation of their IR signal is usually not straightforward. Additionally, electronic structure calculations suggest that, in terms of $\Delta$-splitting, the tilted-bridge geometries resemble the bridging ones\cite{zhang2019identification} and are therefore difficult to resolve experimentally and often not discussed separately.

Previously,\cite{sowa2023exploring} we have shown, using a model acetate-passivated PbS nanocrystal, that IR spectra of small quantum dots can be effectively simulated using molecular dynamics (MD) utilizing machine-learned force fields trained on DFT data.\cite{behler2007generalized,behler2016perspective,unke2021machine,poltavsky2021machine,deringer2019machine} This approach\cite{gastegger2017machine} was demonstrated to circumvent issues associated with alternative methods that make use of analytical force fields,\cite{cosseddu2017force} DFT harmonic mode analysis\cite{zhang2019identification,abuelela2012dft} or exceedingly computationally intensive \textit{ab initio} MD  (AIMD) simulations.\cite{voznyy2016computational} 

In this work, we focus on lead-rich PbS nanocrystals passivated with butyrate and oleate ligands. The results of IR spectroscopy on oleate-passivated PbS (and closely related PbSe) nanocrystals have been repeatedly discussed in the literature.\cite{cass2013chemical,kennehan2020dynamic,kennehan2021influence,peters2019hybrid,peters2019sizing} Following the work of Cass \textit{et al.},\cite{cass2013chemical} it has been generally assumed, primarily due to the broad and highly non-Lorentzian shape of the antisymmetric C-O stretch band, that the observed IR signal indicates a coexistence of chelating and bridging oleate geometries. Our earlier simulations of acetate-passivated nanocrystals suggested that chelating geometries should not be observed in line with prior computational studies of carboxylate-passivated CdSe nanocrystals.\cite{voznyy2011mobile,cosseddu2023ligand}
However, our use of short acetate ligands in our prior work prohibited us from making a direct comparison with experiment.\cite{sowa2023exploring} The primary aim of this work is therefore to resolve this issue by simulating spectra of PbS nanocrystals capped with longer ligands that can also be used experimentally. 

First, based on IR measurements, we demonstrate that oleate and butyrate ligands adopt similar binding geometries on the surface of the nanocrystals, which allows us to use the shorter butyrate ligands in our simulations. A qualitative assignment of the experimental IR peaks is then made by performing DFT harmonic analysis on a single-ligand toy system. Using a machine-learned force field and a dipole algorithm trained on DFT data, we next simulate an IR spectrum of a more realistic model butyrate-passivated PbS nanocrystal.
We find that the experimental spectrum can be successfully reproduced by our simulations in which  no significant amounts of chelating geometries are observed. Instead, the broad and asymmetric spectral bands appear to stem from the wide range of  bridge  and tilted-bridge geometries accessible at room temperature on a disordered nanocrystal-ligand interface. 
These findings highlight the uncertainties associated with empirical interpretation of nanocrystal IR spectra since non-trivial spectral line shapes can be observed even in the presence of only one or two nominal ligand coordination geometries.
Our results also demonstrate the capability of DFT-trained ML algorithms to achieve accuracy and efficiency, thereby enabling effective modeling of vibrational spectra of nanoscopic structures. 


\section{Methods}
\subsection{Experiments}
Oleate-passivated PbS nanocrystals were synthesized using the procedure described by Kessler \textit{et al.}\cite{kessler2018exchange} Butyrate-passivated nanocrystals were obtained by a subsequent solution ligand exchange, see Section S1.1 for details. The average size of the nanocrystals was estimated at 2.8 nm based on the position of their first exciton absorption band,\cite{cademartiri2006size,moreels2009size} see Figure S1. As described in detail in Section S1.3, FTIR spectra of PbS nanocrystals discussed in the main body of this work were measured on films spin-coated on a CaF$_2$ substrate using hexane solutions of freshly synthesized nanocrystals. Spectra of films of aged nanocrystals and films spin-coated from chloroform solutions are discussed in Section S2.2.
\subsection{Simulations}
Unless specified otherwise, all electronic structure calculations were performed using the dispersion-corrected B3LYP-D3 functional,\cite{becke1992density,stephens1994ab,grimme2010consistent} lanl2dz basis set with corresponding effective core potentials\cite{wadt1985ab} on Pb, S, and I, and 6-31G* basis on the remaining light atoms. Calculations on the single-ligand toy system were performed with NWChem 7.0.0\cite{apra2020nwchem} and those on the larger model system with TeraChem 1.9.\cite{Ufimtsev2009b,terachem21} \\
The force field was trained on both energies and forces using the DeePMD framework.\cite{wang2018deepmd,zeng2023deepmd2,zhang2018deep} We used a total of 10,648 geometries and the training was performed for $7\times10^5$ steps with a cut-off of 7 \AA. The performance of the ML force field during an MD simulation is examined in detail in Section S4. We obtain root-mean-square errors on energy and force components of RMSE(energy) = 0.61 meV and RMSE(force) = 81 meV/\AA, well below the recommended thresholds.\cite{wen2022deep} The dipole moment algorithm was also trained using the DeePMD framework on the same set of geometries for $7.5\times10^5$ steps with a cut-off of 8 \AA. Its performance during MD simulations is also shown in Section S4; we find RMSE on dipole vector components of RMSE(dipole) = 0.51 D corresponding to the $R^2$ value of $R^2$ = 0.985.\\
The MD simulations were performed using the Atomic Simulation Environment (ASE)\cite{larsen2017atomic} interfaced with the DeePMD force field using a timestep of 1 fs and, when appropriate, a Langevin thermostat with $\gamma$ = 4.9 ps$^{-1}$. The initial geometry for the NVT simulation was taken from the training set and thermalized for 0.5 ns prior to the NVE simulation which was performed for 3.5 ns. The memspectrum package was used for the Burg’s maximum entropy spectra analysis method.\cite{memspectru}  Frequencies in the calculated spectra were scaled by a factor of 0.96 as appropriate for this choice of functional and basis set.\cite{scott1996harmonic}\\
Additionally, we also trained ML force field and dipole algorithms using DFT data obtained with a smaller (6-31G/lanl2dz) basis set. As we discuss in Section S7, MD simulations based on that level of DFT calculations predict similar distributions of ligand geometries as those based on the larger basis set but this approach yields a poorer agreement between the experimental and simulated vibrational spectrum.\\
All 3D visualisations were rendered with VMD.\cite{humphrey1996vmd}

\section{Results and Discussion}
\subsection{Experimental IR Spectra}
We begin by comparing experimental IR spectra of PbS nanocrystals functionalized with butyrate and (frequently used) oleate ligands and demonstrate that the two systems possess very similar distributions of carboxylate binding geometries. This will allow us to consider only the shorter, and thus computationally more expedient, butyrate ligands in the subsequent simulations.
\begin{figure}
    \centering
    \includegraphics{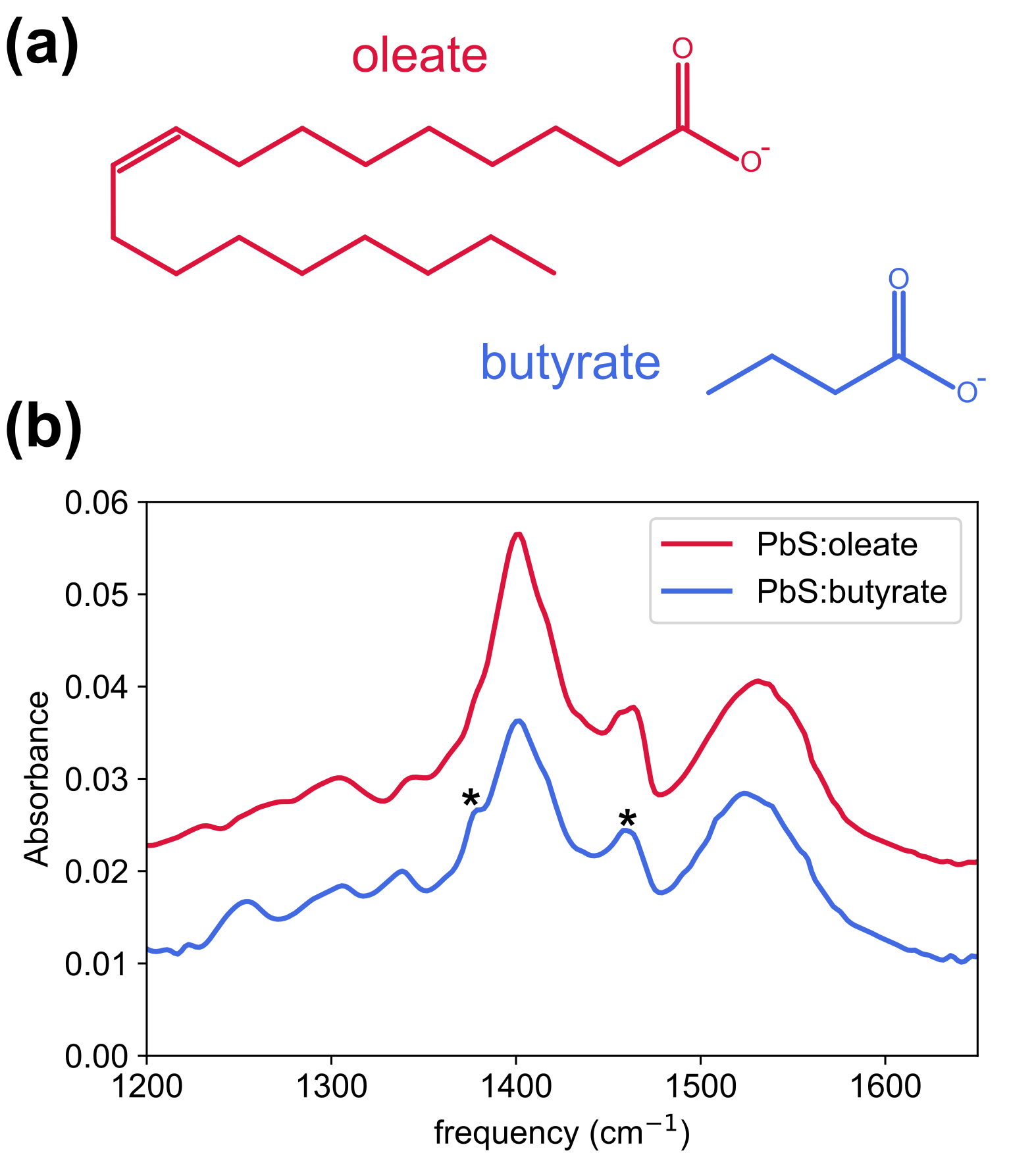}
    \caption{(a) Chemical structures of the considered ligands. (b) IR spectra  of passivated PbS nanocrystals in the region of the ligand carboxyl stretches. For better comparison, the oleate data was offset vertically by 0.015. The 1380 and 1460 cm$^{-1}$ peaks in the butyrate spectrum assigned to impurities are indicated above ($*$).}
    \label{exp}
\end{figure}

We performed FTIR measurements on thin films of oleate- and butyrate-passivated PbS nanocrystals spin-coated onto CaF$_2$ substrates. The used PbS nanocrystals were roughly 2.8 nm in diameter and are therefore expected to posses exclusively lead-rich (111) facets.\cite{choi2013steric,beygi2018surface,kessler2020mapping} As discussed above, they were originally synthesized with oleate ligands and the butyrate ligands were introduced through solution exchange.
The high-frequency parts of the two spectra are shown in Figure S2. We find that, as expected, the exchange of the oleate with the butyrate ligands is accompanied by a virtually complete disappearance of the 3000 cm$^{-1}$ peak corresponding to the alkene C-H bond stretch. Additionally, both spectra confirm lack of significant amounts of hydroxyl or protonated carboxylic acid groups in the sample (\textit{c.f.}~Ref.~\onlinecite{zherebetskyy2014hydroxylation}). 

In Figure \ref{exp}(b), we show the regions of the IR spectra of each film that contain the relevant carboxyl stretch frequencies.  The spectrum of oleate-passivated PbS nanocrystals agrees very well with spectra previously reported in the literature.\cite{kennehan2020dynamic,peters2019sizing,kennehan2021influence,grisorio2016dynamic,giansante2015darker,jeong2012enhanced} We observe a symmetric C-O stretch peak at around 1400 cm$^{-1}$ followed by a band corresponding to C-H bends at around 1450 cm$^{-1}$, and finally a broad and asymmetric peak between 1500-1560 cm$^{-1}$ stemming from the antisymmetric C-O stretches. The progression of bands around and below 1350 cm$^{-1}$ is known to correspond to CH$_2$ wagging and twisting modes.\cite{maroncelli1982nonplanar,peters2019hybrid} The spectrum of the butyrate-passivated nanocrystals features nearly identical line shape and peak positions. The similarity of these spectra indicate that   the oleate and butyrate ligands adopt very similar carboxylate binding geometries on the surface of the nanocrystal.
We note however that, unlike for the oleate-passivated nanocrystals, the spectrum of the butyrate-passivated quantum dots is not stable. As shown in Figure S3, this evolution is associated with an increase in the intensity of the 1380 and 1460 cm$^{-1}$ peaks, indicated in Figure \ref{exp}(b), relative the those corresponding to the carboxyl stretches. We therefore speculate that these two features in the PbS:butyrate spectrum correspond to products of nanocrystal decomposition rather than the butyrate ligands passivating the nanocrystal surface. We note that although the line shape corresponding to these other entities (a set of two peaks at approximately 1380 and 1460 cm$^{-1}$) is typical of most alkane materials, we observe it also in films cast from chloroform suggesting that the solvent \textit{per se} is an unlikely source of these features, see Figure S3.

\subsection{Toy model simulations}
\begin{figure*}
    \centering
    \includegraphics{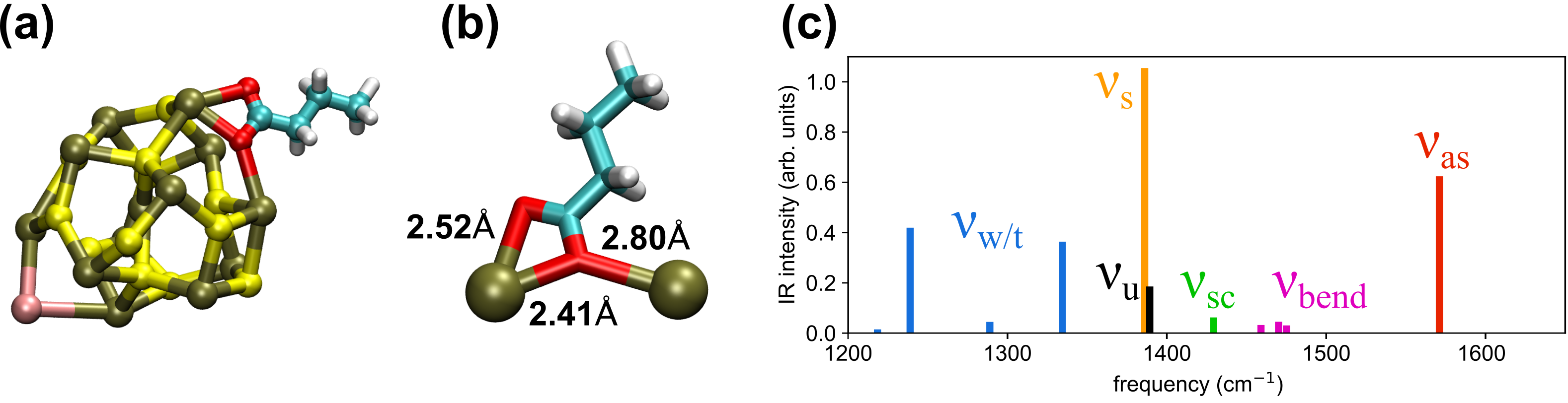}
    \caption{(a) Optimized structure of the toy model system, Pb$_{13}$S$_{12}$I:butyrate nanocrystal. (b) The optimized geometry of the butyrate ligand from (a). (c) Harmonic IR spectrum  of the toy system with modes identified by inspection. See text for description of the vibrational mode labels.}
    \label{toy}
\end{figure*}
We next move on to the computational considerations. It is helpful to begin by studying a toy model system comprising a single butyrate ligand attached to a Pb$_{13}$S$_{12}$I nanocrystal (the iodine atom was added to balance the charge/unpaired electrons without affecting the vibrational spectrum), see Figure \ref{toy}(a). Following the geometry optimization, the ligand relaxes to a ‘tilted-bridge’-type geometry with Pb-O bond lengths as indicated in panel (b). The IR spectrum calculated for this geometry using harmonic analysis is shown in Figure 3(c) and, very broadly, agrees with the experimental spectra of butyrate-capped nanocrystals and confirms the mode assignment discussed above. The model features a series of modes between 1200 and 1350 cm$^{-1}$ that can be assigned to CH$_2$ wagging and twisting modes, $\nu_\mathrm{w/t}$. Around 1390 cm$^{-1}$ we find the intense symmetric C-O stretch, $\nu_\mathrm{s}$, neighboring the much less intense umbrella mode of the methyl group, $\nu_\mathrm{u}$, which depending on the conformation, can have a frequency above or below that of the symmetric C-O stretch. The mode around 1570 cm$^{-1}$ corresponds to the antisymmetric C-O mode, $\nu_\mathrm{as}$. The above calculation also predicts a scissoring mode of the methylene group adjacent to the carboxylate, $\nu_\mathrm{sc}$,  at around 1430 cm$^{-1}$. 
We note however that the calculated intensity of this mode is much weaker than that of the 1460 cm$^{-1}$ peak observed experimentally in the butyrate spectrum which we assigned to an impurity stemming from nanocrystal degradation. The remaining C-H bends, $\nu_\mathrm{bend}$, are similarly found to possess almost negligible IR intensities.

In Section S3, we discuss results of harmonic analyses performed on ten additional geometries obtained by optimizing different initial configurations, all of which relaxed into either bridging or tilted-bridging geometries. While the spectra remain qualitatively similar, we observe variations in IR intensities, vibrational frequencies, and the splitting $\Delta$ between the symmetric and antisymmetric C-O stretch frequencies. As shown in Figure S4, we find that the $\nu_\mathrm{as}$ frequencies span a range of over 60 cm$^{-1}$ while $\Delta$ varies between approximately 120-190 cm$^{-1}$.
\subsection{ML-MD simulations}
Although the calculations performed on the toy system above appear to confirm the conventional mode assignment, they yield no information regarding the IR line shape or the underlying distribution of ligand geometries. Furthermore, harmonic analysis does not capture possible anharmonic effects and, as discussed above, its results can be strongly conformation-dependent. Finally, preferred ligand binding geometries may qualitatively differ on well-defined crystal facets and may change due to ligand-ligand interactions as the ligand density on the surface is increased.

\begin{figure}
    \centering
    \includegraphics{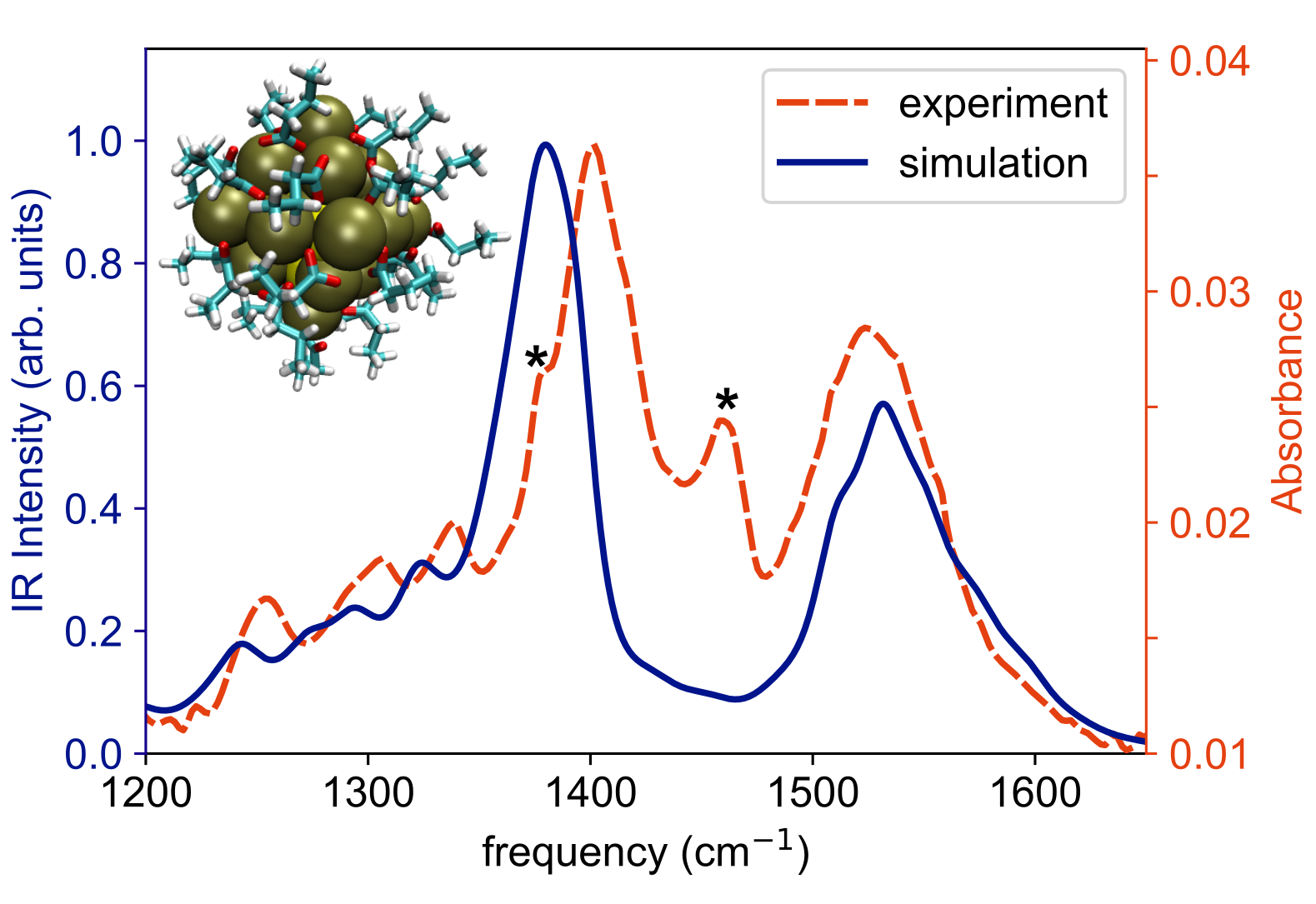}
    \caption{The experimental and simulated IR spectra of butyrate-passivated nanocrystals. Asterisks indicate peaks assigned to impurities. The simulated spectrum was obtained using Burg’s maximum entropy spectral analysis method;\cite{burg1975maximum} its comparison with the numerical Fourier transform result is shown in Figure S7.  Inset: The Pb$_{19}$S$_6$(butyrate)$_{26}$ nanocrystal considered in the simulations.}
    \label{MLMD}
\end{figure}

To address these issues, we proceed to consider a more realistic model system, shown schematically in Figure \ref{MLMD}, which comprises a small octahedral non-stoichiometric Pb$_{19}$S$_6$ nanocrystal\cite{choi2013steric} passivated by 26 butyrate ligands, yielding the total system closed-shell and charge-neutral. Although the considered nanocrystal is significantly smaller than the ones used experimentally, in analogy to the experimentally used quantum dots,\cite{choi2013steric,beygi2018surface,kessler2020mapping} it possesses exclusively lead-rich (111) facets and should therefore support ligand binding geometries similar to those seen experimentally.

The IR spectrum for this system is obtained by considering the time-dependence of the total dipole moment during a constant-energy MD simulation.\cite{thomas2013computing} To perform the necessary simulations, we trained a ML force field algorithm on DFT energies and forces; see Methods for details. Additionally, we trained a ML algorithm to predict the total (DFT) dipole moment based on the atomic coordinates. The dipole vector is then calculated during an NVE simulation performed on a thermalized initial state (at $T = 300$ K). Finally, the IR signal is calculated from the dipole-dipole correlation function,\cite{iftimie2005ab} 
\begin{equation}
    I(\omega) \propto  \int_{-\infty}^\infty \mathrm{d}t \: \langle \dot{\mu}(\tau) \dot{\mu}(t-\tau) \rangle_\tau \: e^{\mathrm{i}\omega t},
\end{equation}
where $\dot{\mu}(t)$ is the derivative of the dipole moment at time $t$ and $\langle \cdot \rangle_\tau $ denotes an ensemble average over $\tau$. We note that the expression above already contains the harmonic quantum correction factor.\cite{bader1994quantum}

The relevant region of the resulting spectrum is shown in Figure \ref{MLMD} (the spectrum over the entire frequency range is shown in Figure S6). It features two intense peaks corresponding to the symmetric and antisymmetric C-O stretches at around 1390 and 1520-1550 cm$^{-1}$, respectively. Their relative positions, heights, and widths agree well with the experiment, suggesting that the simulations correctly capture the dynamics of the carboxylate ligands. The simulations also very well capture the collective CH$_2$ vibrations (progression of bands between  roughly 1200 and 1350 cm$^{-1}$).
We note however that peaks previously assigned to impurities (around 1380 and 1460 cm$^{-1}$) are absent in the simulated spectrum.
As we will demonstrate below, there exist C-H bending modes of frequencies between 1400 and 1500 cm$^{-1}$ in the underlying nuclear dynamics. In agreement with the single-ligand calculations, however, their low intensities yield them virtually silent. As we demonstrate in Section S6, the very low intensities of the scissoring modes and the remaining C-H bends are independent of the used electronic structure method, further confirming our assignment of the 1460 cm$^{-1}$ peak to impurities rather than a C-H bend within the surface butyrate ligands. 



\begin{figure}[b]
    \centering
    \includegraphics{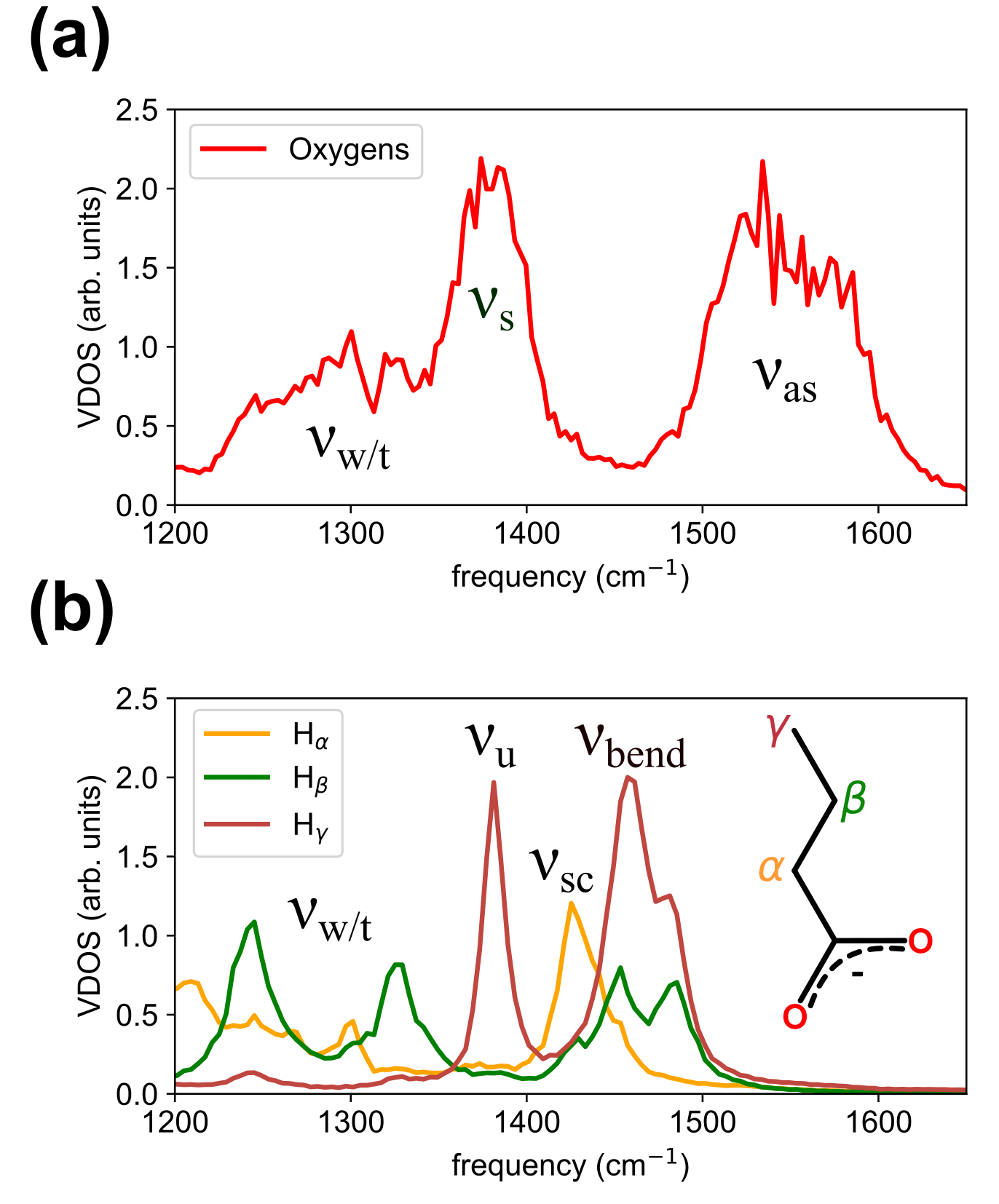}
    \caption{(a, b) VDOS for (a) oxygen and (b) hydrogen atoms calculated from a 100 ps NVE simulation.}
    \label{vdos}
\end{figure}

To further understand the simulated spectrum, we next calculate the vibrational densities of states (VDOS) for chemically distinct atom types from the individual atomic velocity correlation functions,
\begin{equation}
    g_X(\omega) \propto   \int_{-\infty}^\infty \mathrm{d}t \sum_{n \in X} \langle v_n(\tau) v_n(t-\tau) \rangle_\tau \: e^{\mathrm{i}\omega t},
\end{equation}
where $v_n(t)$ is the velocity of atom $n$ at time $t$, and the sum runs over all chemically equivalent atoms of type $X$. 
VDOS for the oxygen and hydrogen atoms are plotted in Figures \ref{vdos}(a) and \ref{vdos}(b), respectively.
In the oxygen VDOS, as expected, we find two bands corresponding to the symmetric, $\nu_\mathrm{s}$, and antisymmetric C-O stretches, $\nu_\mathrm{as}$. In the case of the hydrogen atoms, in agreement with the single-ligand toy system, we find evidence of the methyl group umbrella modes, $\nu_\mathrm{u}$, overlapping with the symmetric C-O stretches. We can also confirm our assignment of the bands between 1200 and 1350 cm$^{-1}$ as predominantly CH$_2$ bending, $\nu_\mathrm{w/t}$, although we note some oxygen contribution to vibrations in this frequency range. Finally, in the VDOS of the hydrogen atoms, there exists a clear peak at a frequency of roughly 1430 cm$^{-1}$ that corresponds to the previously discussed scissoring mode, $\nu_\mathrm{sc}$, as well as peaks corresponding to the remaining C-H bends, $\nu_\mathrm{bend}$, \textit{c.f.}~Figure \ref{toy}(c). However, as discussed above and in Section S6, these modes are almost silent and thus virtually absent in the simulated IR spectrum.


Lastly, we investigate the geometries adopted by the butyrate ligands during the NVE simulation (and therefore directly corresponding to the observed IR spectrum). We use a method introduced in our previous work.\cite{sowa2023exploring} Briefly, for each ligand, we identify its primary lead atom, Pb$^*$, as the lead atom closest to the average position of the two oxygen atoms. The two O-Pb$^*$ distances are then defined as $r_1$ and $r_2$, where $r_2 > r_1$. The shortest Pb-O distance to any other lead atom is defined as $r_3$. In the case of the ligand shown in Figure 3(b), for instance, we find $r_1$ = 2.41, $r_2$ = 2.52 and $r_3$ = 2.80 \AA. The distance $r_1$ will generally correspond to that of a Pb-O (primarily ionic) bond. In the case of a chelating geometry, $r_1\sim r_2$ and both are significantly shorter than $r_3$. For a bridging ligand, on the other hand, $r_1$ and $r_3$ distances should be comparable and shorter than $r_2$. Finally, for a tilted-bridge, $r_1\sim r_2\sim r_3$ while for a unidentate geometry both $r_2$ and $r_3$ should be significantly longer than $r_1$.
\begin{figure*}
    \centering
    \includegraphics{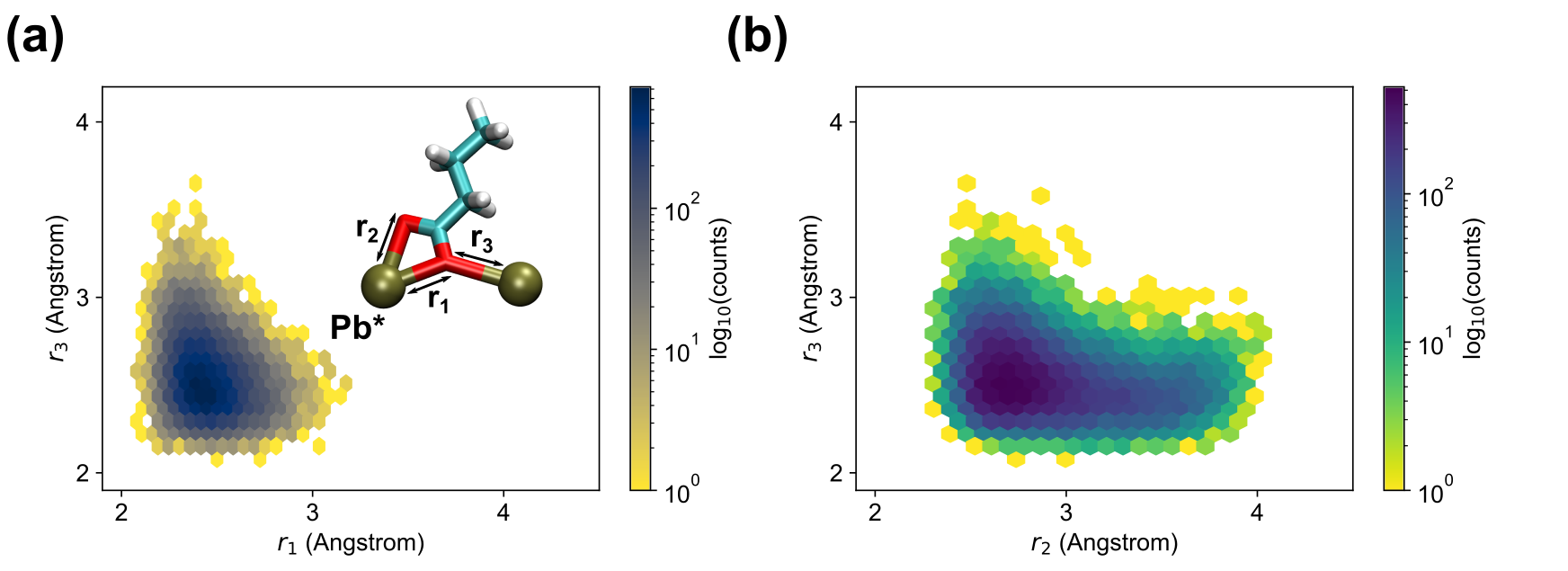}
    \caption{Histograms showing (a) $r_1$ vs $r_3$ and (b) $r_2$ vs $r_3$ values adopted during the NVE simulation used to obtain the IR spectrum in Figure \ref{MLMD}.}
    \label{r123}
\end{figure*}

The results of the above analysis, the $r_1$ vs $r_3$ and $r_2$ vs $r_3$ histogram plots, are shown in Figure \ref{r123}(a) and (b). They reveal a very wide range of $r_1$, $r_2$, and $r_3$ values adopted by the ligands at room temperature in our simulations. Nonetheless, we observe a single maximum in both histograms, indicating that the tilted-bridge geometries dominate although some amount of bridging ligands can also be found. This is in broad agreement with our earlier work on acetate-passivated nanocrystals in which a different functional as well as a smaller basis set was used, suggesting that this result is largely independent of the chosen electronic structure method. Crucially, although the peak corresponding to the antisymmetric C-O stretches possesses a highly non-Lorentzian structure, we find no evidence for existence of significant numbers of chelating geometries or, in fact, for a bimodal distribution of ligand geometries of any type. Instead, this asymmetric and seemingly structured spectrum appears to arise due to an inherently broad and asymmetric distribution of (largely) tilted-bridge geometries. The highly disordered nature of the nanocrystal-ligand interface implies that the experimental C-O stretch bands cannot be interpreted simply as a sum of peaks corresponding to distinct coordination types. This is especially true given that, as shown in Figure S4, the conformation of carboxylate ligands on the nanocrystal surface affects not only the C-O stretch frequencies but also their corresponding intensities. 

\section{Conclusions}
In summary, we have studied the experimental vibrational spectra of oleate and butyrate-passivated PbS nanocrystals and demonstrated that these two ligands possess very similar binding geometries. This allowed us to study the computationally more expedient butyrate ligands in lieu of the oleate ligands customarily used experimentally to enhance the stability of colloidal nanocrystals. We first considered a single-ligand toy system, which allowed for a straightforward mode assignment through a harmonic mode analysis. We then investigated a more realistic model comprising a small octahedral PbS nanocrystal passivated by twenty-six butyrate ligands. The corresponding IR spectrum was obtained from an MD simulation performed with a machine-learned force field trained on DFT data. We obtained good agreement with the experimental spectrum and therefore confirmed that ML-AIMD simulations constitute an effective tool in modelling the IR response of nanoscopic systems.

Our work also resolves the apparent inconsistency between the conventional interpretation of the IR spectroscopy of carboxylate-passivated PbS nanocrystals\cite{cass2013chemical,kennehan2020dynamic,kennehan2021influence,peters2019hybrid,peters2019sizing} and previous computational studies on these systems.\cite{sowa2023exploring} The former postulated a coexistence of bridging and chelating ligand geometries while the latter suggested that chelating geometries should not be observed. Our simulations indicate that the experimental IR line shape can result from a broad distribution of tilted-bridge geometries. Importantly, this work illustrates limitations of interpreting nanocrystal IR spectra based on empirical rules derived from studies of metal complexes since the very wide range of conformations accessible to carboxylate ligands can result in broad and non-Lorentzian IR line shapes even if few nominal coordination types are present on a nanocrystal's surface. 

\section*{Acknowledgement}
This work was supported by the Center for Adopting Flaws as Features, an NSF Center for Chemical Innovation supported by grant CHE-2413590. We also acknowledge the Big-Data Private-Cloud Research Cyberinfrastructure MRI award funded by the NSF under grant CNS-1338099 and by Rice University’s Center for Research Computing (CRC). 
DMC thanks NSF for support under the Graduate Research Fellowship Program, grant DGE-1610403. STR and DMC acknowledge additional materials support from the Welch Foundation under grant F-1885.
AM and BGL gratefully acknowledge funding from the Institute for Advanced Computational Science.  This work used Expanse GPU at the San Diego Supercomputer Center (SDSC) through allocation CHE140101 from the Advanced Cyberinfrastructure Coordination Ecosystem: Services and Support (ACCESS) program, which is supported by National Science Foundation grants No. 2138259, 2138286, 2138307, 2137603, and 2138296.

\section*{Supplementary Information}
Details of experimental procedure and additional experimental results; details of the ML training procedure; additional toy-model results; additional ML-MD results; discussion regarding the mode intensity; results obtained using the smaller-basis-set force field.

\end{document}